\documentclass[fleqn,usenatbib]{mnras}

\usepackage{newtxtext,newtxmath}

\usepackage[T1]{fontenc}

\DeclareRobustCommand{\VAN}[3]{#2}
\let\VANthebibliography\thebibliography
\def\thebibliography{\DeclareRobustCommand{\VAN}[3]{##3}\VANthebibliography}

\usepackage{graphicx}	
\usepackage{amsmath}	
\usepackage{fontawesome}
\usepackage{multirow}

\newcommand{\grad}{\ensuremath{\vec{\nabla}}}
\newcommand{\adotoa}{\ensuremath{{\cal H}}}

\newcommand{\Pb}{\ensuremath{\bar{P}}}
\newcommand{\rhob}{\ensuremath{\bar{\rho}}}

\newcommand{\phib}{\ensuremath{\bar{\phi}}}
\newcommand{\Zb}{\ensuremath{\bar{Z}}}

\newcommand{\CLASS}{\textsc{class}}

\newcommand{\be}{\begin{equation}}
\newcommand{\ee}{\end{equation}}
\newcommand{\bea}{\begin{eqnarray}}
\newcommand{\eea}{\end{eqnarray}}

\title[KIDS-1000 + \textsc{CosmoPower}: Interacting Dark Energy]{KiDS-1000 Cosmology: machine learning - accelerated constraints on Interacting Dark Energy with \textsc{CosmoPower}}

\author[A. Spurio Mancini \& A. Pourtsidou]{
A. Spurio Mancini$^{1}$\thanks{E-mail: a.spuriomancini@ucl.ac.uk},
A. Pourtsidou$^{2,3,4}$
\\
$^{1}$Mullard Space Science Laboratory, University College London, Holmbury St. Mary, Dorking, Surrey, RH5 6NT, UK\\
$^{2}$Institute for Astronomy, The University of Edinburgh, Royal Observatory, Edinburgh EH9 3HJ, UK\\
$^{3}$Department of Physics \& Astronomy, University of the Western Cape, Cape Town 7535, South Africa\\
$^{4}$School of Physical and Chemical Sciences, Queen Mary University of London, Mile End Road, London E1 4NS, UK
}
\date{Accepted XXX. Received YYY; in original form ZZZ}

\pubyear{2021}

\begin{document}
\label{firstpage}
\pagerange{\pageref{firstpage}--\pageref{lastpage}}
\maketitle

\begin{abstract}
We derive constraints on a coupled quintessence model with pure momentum exchange from the public $\sim$1000 deg$^2$ cosmic shear measurements from the Kilo-Degree Survey and the \textit{Planck} 2018 Cosmic Microwave Background data. We compare this model with $\Lambda$CDM and find similar $\chi^2$ and log-evidence values. We accelerate parameter estimation by sourcing cosmological power spectra from the neural network emulator \textsc{CosmoPower}. We highlight the necessity of such emulator-based approaches to reduce the computational runtime of future similar analyses, particularly from Stage IV surveys. As an example, we present MCMC forecasts on the same coupled quintessence model for a \textit{Euclid}-like survey, revealing degeneracies between the coupled quintessence parameters and the baryonic feedback and intrinsic alignment parameters, but also highlighting the large increase in constraining power Stage IV surveys will achieve. The contours are obtained in a few hours with \textsc{CosmoPower}, as opposed to the few months required with a Boltzmann code. 
\end{abstract}



\section{INTRODUCTION}
Current and forthcoming large-scale structure (LSS) surveys such as the Dark Energy Survey\footnote{\url{www.darkenergysurvey.org}}, ESA's  \textit{Euclid} satellite mission\footnote{\url{www.euclid-ec.org}}, and  the Vera C. Rubin Observatory’s Legacy Survey of Space and Time (VRO/LSST)\footnote{\url{https://www.lsst.org/}}, 
are aiming to probe the nature of the dark sector (dark energy and dark matter) by performing high precision galaxy clustering and weak gravitational lensing measurements. The standard model of cosmology, $\Lambda$CDM, is currently providing the best fit to a suite of data from Cosmic Microwave Background (CMB) and LSS experiments (e.g. \citealt{Aghanim:2018eyx, Anderson_2012, Song:2015oza, Beutler_2016, Troster:2019ean, Alam:2020sor, Abbott:2021bzy, Heymans_2021}). $\Lambda$CDM assumes that dark energy is a cosmological constant, $\Lambda$, and that General Relativity describes gravity on all scales. It also assumes that dark energy and dark matter are non-interacting (uncoupled). LSS surveys are aiming to constrain exotic dark energy and modified gravity models (for reviews see e.g. \citealt{Copeland:2006wr, Clifton:2011jh}).

In this work we focus on constraining interacting dark energy (IDE) in the form of a scalar field $\phi$ (quintessence) explicitly coupled to cold dark matter (CDM). IDE models have been widely studied and have gained popularity as potential alternatives to $\Lambda$CDM \citep{Amendola:1999er,Pourtsidou:2013nha, Tamanini:2015iia, DiValentino:2019jae, Lucca:2021dxo}. Here we study a sub-class of models that only exhibit momentum exchange between dark energy and dark matter \citep{Simpson:2010vh, Pourtsidou:2013nha, Baldi:2014ica, Baldi:2016zom,  Chamings:2019kcl,Amendola:2020ldb,Kase:2020hst}. This allows them to fit CMB, supernovae, and baryon acoustic oscillation data very well \citep{Pourtsidou:2016ico, Linton:2021cgd}, but they have not been tested yet with weak lensing data marginalising over baryonic feedback effects. 

Baryonic and dark matter nonlinear effects become particularly important in weak lensing studies with Stage IV surveys like \textit{Euclid} and VRO/LSST, as they dominate the small, nonlinear scales with the most constraining power \citep{Schneider:2019snl,Schneider:2019xpf,Martinelli:2020yto}. At the same time, the computational requirements for accurate parameter estimation are becoming very expensive. A typical Markov Chain Monte Carlo (MCMC) requires $> 10^4$ evaluations of the theoretical model under consideration, with the runtime being dominated by the computation of cosmological power spectra with Boltzmann codes such as \textsc{CAMB} \citep{Lewis:1999bs} or \textsc{CLASS} \citep{Lesgourgues:2011re, Blas:2011rf}. This has led to the development of fast power spectra emulators \citep[e.g.][]{Arico:2021izc, mootoovaloo2021kernelbased, SpurioMancini:2021ppk} to accelerate the inference pipeline by replacing the Boltzmann code at each likelihood evaluation.     

\section{MODEL}
\label{sec:IDEmodel}

The model we study belongs to the pure momentum transfer class of theories constructed in~\citet{Pourtsidou:2013nha, Skordis:2015yra}. Its main feature is that no coupling appears at the background level, regarding the fluid equations. This is in contrast to the most commonly considered coupled quintessence models, but it is also what makes this model able to fit data for a wide range of the coupling parameter $\beta$ \citep{Pourtsidou:2016ico}. 
In addition, the energy-conservation equation remains uncoupled even at the linear perturbations level. Therefore, the model provides for a pure momentum-transfer coupling at the level of linear perturbations. 

Following \citet{Pourtsidou:2016ico} we are going to concentrate on the case where the action for the scalar field $\phi$ is written as 
\bea
 \nonumber
S_{\phi}&=&
\int dt \, d^3x \, a^3 \left[
    \frac{1}{2}(1 - 2 \beta)  \dot{\phi}^2
 - \frac{1}{2} |\grad\phi|^2
- V(\phi)
\right] \, .
\label{eq:action}
\eea
The model is physically acceptable for $\beta < \frac{1}{2}$. For $\beta \rightarrow 1/2$
there is a strong coupling pathology, while for $\beta>1/2$ there is a ghost in the theory since the kinetic term becomes negative.

\subsection{Background Evolution}
Assuming a flat Friedmann-Lema\^itre-Robertson-Walker (FLRW) Universe, the background energy density and pressure for quintessence are~\citep{Pourtsidou:2013nha}
\bea
\rhob_{\phi}=\left(\frac{1}{2}-\beta\right)\frac{\dot{\phib}^2}{a^2}+V(\phi);  \;\;
\Pb_{\phi}=\left(\frac{1}{2}-\beta\right)\frac{\dot{\phib}^2}{a^2}-V(\phi)\, ,
\eea
and the energy conservation equations are the same as in uncoupled quintessence:
\bea
\dot{\rhob}_\phi +  3{\cal H}(\rhob_\phi+\Pb_\phi)=0; \;\; \dot{\rhob}_c + 3{\cal H}\rhob_c=0 \, .
\eea

\subsection{Linear Perturbations}
In order to study the observational effects of the coupled models on the Cosmic Microwave Background and Large-Scale Structure (LSS), we need to consider linear perturbations around the FLRW background. The density contrast $\delta_c \equiv \delta \rho_c / \rhob_c$ obeys the standard evolution equation
\bea
\label{eq:deltac}
\dot{\delta}_c = -k^2 \theta_c - \frac{1}{2}\dot{h}.
\eea 
The momentum-transfer equation depends on the coupling parameter, $\beta$, and is given by
\bea
 \dot{\theta}_c = - \adotoa  \theta_c + 
 \frac{( 6 \adotoa \beta \Zb + 2\beta \dot{\Zb} )  \varphi +  2\beta \Zb  \dot{\varphi}}{a \left(\rhob_c  - 2 \beta \Zb^2 \right)} \, ,
 \label{dP_type3_pert}
\eea
where $\phi = \bar{\phi} + \varphi$, and $\bar{Z}= - \dot{\bar{\phi}}/a$.
We implemented the above equations in \CLASS{} \citep{Lesgourgues:2011re,Blas:2011rf} in order to compute the CMB temperature and matter power spectra, following the previous implementation in \citet{Pourtsidou:2016ico}. We fix the quintessence potential $V(\phi)$ to be the widely used single exponential form (1EXP)
\bea
V(\phi)=V_0 e^{-\lambda \phi}.
\eea 
Our initial conditions for the quintessence field are $\phi_i=10^{-4}, \dot{\phi}_i=0$. However, the same cosmological evolution is expected for a wide range of initial conditions~\citep{Copeland:2006wr}.

\subsection{Nonlinear effects}

To exploit the constraining power of forthcoming large-scale structure datasets on IDE models it is crucial to accurately model nonlinear effects. N-body simulations for momentum exchange in the dark sector have been performed in \citet{Baldi:2014ica, Baldi:2016zom}, based on the elastic scattering model presented in \citet{Simpson:2010vh}. However, for the model considered here there is no available non-linear prescription or N-body data. In our analysis we employ the nonlinear correction implemented in \textsc{HMcode} \citep{Mead:2020vgs}, which includes modelling of baryonic feedback effects. We remark that this prescription is based on the $\Lambda$CDM model. Following \citet{SpurioMancini:2019rxy}, we justify this choice with the expected limited impact of different nonlinear prescriptions on cosmological constraints from the KiDS dataset, given the range of scales probed. However, this approach will need to be modified for applications to future surveys, whose dark energy constraints will strongly depend on the nonlinear prescription adopted. We will return to this issue in \autoref{sec:conclusions} in the context of IDE models, and discuss ways forward.

\section{Data and methods}
\begin{table*}
\resizebox{\textwidth}{!}{%
    \begin{tabular}{|c|ccc|ccc|ccc|ccc|}
                          & \multicolumn{3}{c}{\textbf{Band Powers}}                       & \multicolumn{3}{c}{\textbf{COSEBIs}}                           & \multicolumn{3}{c}{\textbf{2PCFs}} & \multicolumn{3}{c}{\textbf{\textit{Planck}}}\\
                          \hline
                          & $\Lambda$CDM              & IDE                       & IDE ($\lambda=1$) & $\Lambda$CDM                & IDE                       & IDE ($\lambda=1$) & $\Lambda$CDM                & IDE & IDE ($\lambda=1$) & $\Lambda$CDM & IDE & IDE ($\lambda=1$)\\
                          \hline 
                          \hline 
    $\Omega_{\mathrm{m}}$ & $0.341_{-0.076}^{+0.057}$ & $0.342_{-0.083}^{+0.065}$ & $0.343_{-0.084}^{+0.049}$ & $0.314_{-0.083}^{+0.057}$ & $0.315_{-0.086}^{+0.067}$ & $0.318_{-0.087}^{+0.049}$ & $0.269_{-0.055}^{+0.030}$  & $0.272_{-0.059}^{+0.034}$  & $0.270_{-0.056}^{+0.027}$ & $0.320_{-0.009}^{+0.009}$ & $0.318_{-0.009}^{+0.009}$ & $0.335_{-0.009}^{+0.009}$ \\
    \hline
    $\sigma_8$            & $0.714_{-0.105}^{+0.083}$ & $0.714_{-0.107}^{+0.069}$ & $0.722_{-0.106}^{+0.091}$ & $0.743_{-0.095}^{+0.091}$ & $0.745_{-0.090}^{+0.094}$ & $0.751_{-0.114}^{+0.091}$ & $0.816_{-0.068}^{+0.082}$  & $0.812_{-0.068}^{+0.080}$ & $0.830_{-0.073}^{+0.082}$ & $0.813_{-0.008}^{+0.008}$ & $0.814_{-0.008}^{+0.008}$ & $0.790_{-0.008}^{+0.008}$\\
    \hline
    $S_{8}$               & $0.749_{-0.023}^{+0.024}$ & $0.751_{-0.023}^{+0.025}$ & $0.760_{-0.029}^{+0.026}$ & $0.747_{-0.019}^{+0.023}$ & $0.751_{-0.019}^{+0.024}$ & $0.760_{-0.028}^{+0.025}$ & $0.765_{-0.019}^{+0.020}$ & $0.765_{-0.019}^{+0.020}$ & $0.780_{-0.029}^{+0.023}$ & $0.839_{-0.017}^{+0.018}$ & $0.839_{-0.017}^{+0.017}$ & $0.835_{-0.015}^{+0.018}$\\
    \hline
    $\chi^2$              & 148.0036                  & 148.2647                  & 148.7240                  & 77.9787                   & 77.5061                   & 78.4702                   & 255.4080                   & 256.4388                 & 254.7876              & 980.7286                  & 980.7316                  & 980.5730 \\
    \hline
    $\log \frac{\mathcal{Z}_{\mathrm{IDE}}}{\mathcal{Z}_{\Lambda\mathrm{CDM}}}$   & & $-0.055 \pm 0.144$      & $-0.240 \pm 0.140$        &  &  $0.136 \pm 0.146$      & $-0.048 \pm 0.148$       & & $-0.048 \pm 0.183$        & $0.151 \pm 0.184$     &      &    $0.980 \pm 0.277$ & $0.402 \pm 0.279$
    \end{tabular}
    }
    \caption{Mean and marginalised 68 per cent contours on key weak lensing parameters. We also report the $\chi^2$ and log-Bayes factors $\log \frac{\mathcal{Z}_{\mathrm{IDE}}}{\mathcal{Z}_{\Lambda\mathrm{CDM}}}$ values. For the LSS probes the log-Bayes factors are always smaller than 0.5 in absolute value; following \citet{Jeffreys1961}, these values indicate that neither of the two models is clearly favoured with respect to the other. The \textit{Planck} value indicates the CMB data favour the IDE model, although not in a substantial way.}
    \label{tab:posterior}
\end{table*}
We consider the same $\sim$ 1000 deg$^2$ cosmic shear data from the KiDS survey (KiDS-1000) used in the recent analysis of \citet[][A21 in the following]{Asgari21}. Photometric redshift distributions, shear measurements and data modelling are the same presented in the KiDS-1000 papers (\citealt{Hildebrandt21, Giblin21, Joachimi21}). As in A21, we consider three types of cosmic shear summary statistics, namely band powers \citep{Schneider_2002}, Complete Orthogonal Sets of E/B-Integrals \citep[COSEBIs,][]{Schneider_2010}, and two-point real space correlation functions (2PCFs).

We sample the posterior distribution using the Python wrapper \textsc{PyMultiNest} \citep{Buchner14} of the nested sampler \textsc{MultiNest} \citep{Feroz_2008}, as embedded in \textsc{MontePython} \citep{brinckmann2018montepython}. We compare constraints obtained running the KiDS-1000 inference pipeline (for band powers, COSEBIs and 2PCFs) and the \textit{Planck} 2018 TTTEEE+lowE joint polarisation and temperature analysis \citep{Planck18like}. We use \textsc{CosmoPower} \citep[\href{https://github.com/alessiospuriomancini/cosmopower}{\faicon{github}}]{SpurioMancini:2021ppk} to replace the Boltzmann software \textsc{Class} in the computation of the matter and CMB power spectra. All contours shown in \autoref{sec:results_kidsplanck} have been obtained with \textsc{CosmoPower}. An accuracy comparison between \textsc{CosmoPower} and \textsc{Class} contours is reported in \autoref{sec:results_euclid}, where forecast contours are reported for a Stage IV survey configuration, obtained sourcing power spectra from \textsc{CosmoPower} and \textsc{Class}. The technical details of the neural network emulators are unchanged with respect to those described in \citet{SpurioMancini:2021ppk}.

Prior distributions for the sampled parameters are the same used in A21, with the addition of two uniform distributions for the IDE parameters $\beta \sim \mathcal{U}[-0.5, 0.5]$ and $\mathrm{log} \, \lambda \sim \mathcal{U}[-3, 0.32]$. We consider a uniform prior on $\mathrm{log} \, \lambda$ to account for the fact that $\lambda$ is not a dimensionless quantity \citep{Mackay2003}. Choosing uninformative priors is crucial to avoid obtaining constraints driven by the prior assumptions \citep{Simpson2017, Heavens2018}. We also report results obtained fixing $\lambda$ to 1 \citep{Copeland:1997et}. The covariance matrix is the same used in A21. Its analytical computation in $\Lambda$CDM is described in \citet{Joachimi21}; we do not recompute the covariance in the IDE scenario, because similarly to \citet{SpurioMancini:2019rxy} we expect only a weak dependence of the theoretical predictions for the observables on the IDE parameters, verified by the weak constraints obtained on these parameters (see \autoref{sec:results_kidsplanck}).  

\section{Results}\label{sec:results}

\subsection{Constraints from KiDS-1000 and \textit{Planck}}\label{sec:results_kidsplanck}
\begin{figure}
    \centering
    \includegraphics[scale=0.35]{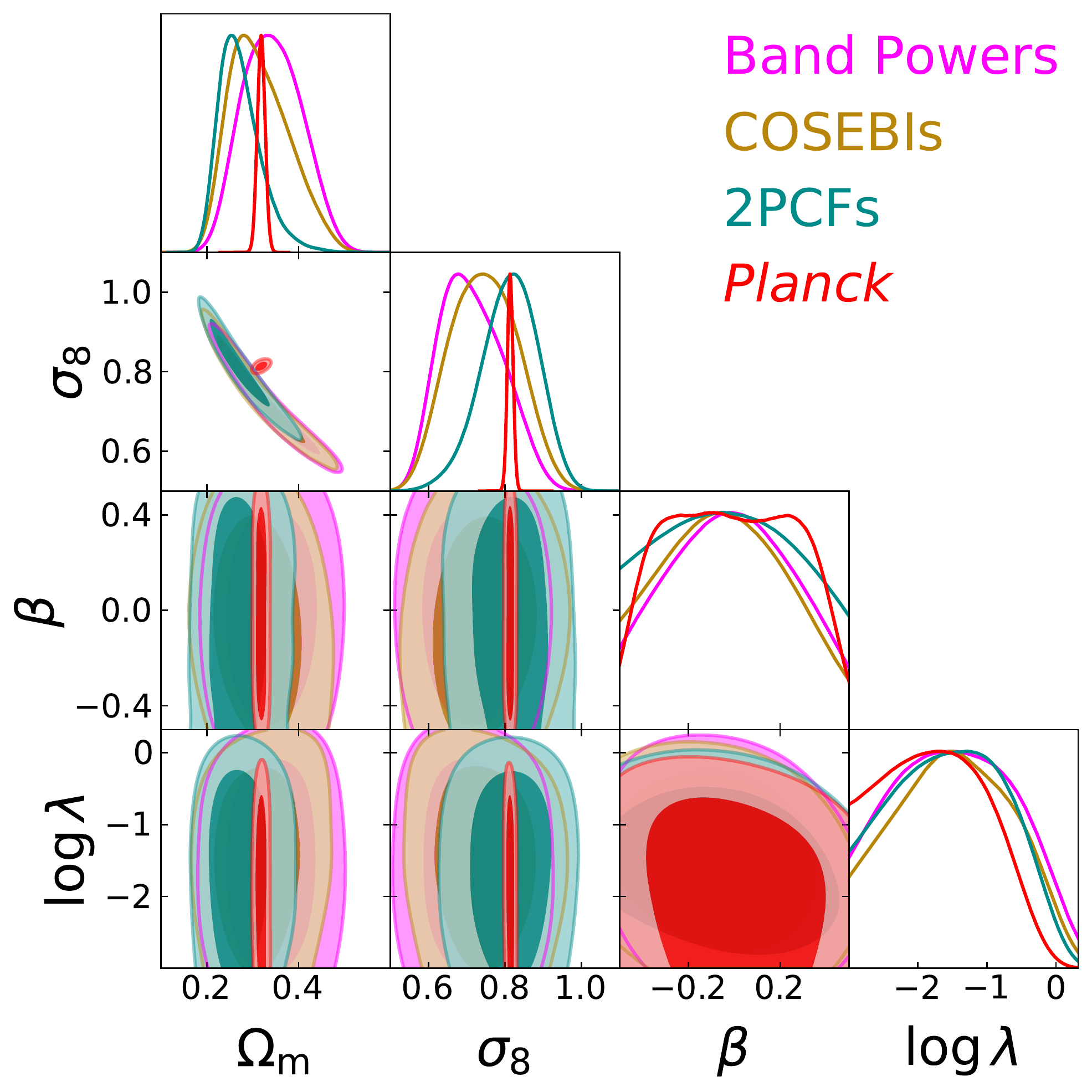}
    \caption{68 and 95 per cent marginalised contours for key weak lensing parameters $\Omega_{\mathrm{m}}, \sigma_8, S_8$ and the IDE parameters $\beta, \lambda$. Contours for band powers, COSEBIs and two-point correlation functions are shown in \textit{magenta, brown} and \textit{cyan}, respectively, while \textit{Planck} contours in \textit{red}.}
    \label{fig:constraints_beta}
\end{figure}

\autoref{fig:constraints_beta} shows a comparison of marginalised 68 and 95 per cent contours of the posterior distribution for the key parameters $\Omega_{\mathrm{m}}, \sigma_8$ and $S_8 = \sigma_8 \sqrt{\Omega_{\mathrm{m}}/0.3}$, as well as for the IDE parameters $\beta, \lambda$. As expected, the latter are unconstrained: differences in the matter power spectrum predictions for IDE models with respect to $\Lambda$CDM are mostly significant at highly nonlinear scales, only very mildly probed by the KiDS-1000 data. The \textit{Planck} likelihood does not constrain $\beta$ and $\lambda$ either, in agreement with the fact that the CMB power spectra are essentially insensitive to these parameters, except on very large, cosmic variance - dominated scales (\citealt{Pourtsidou:2016ico}).

\begin{figure*}
    \centering
    \includegraphics[width=0.3\textwidth]{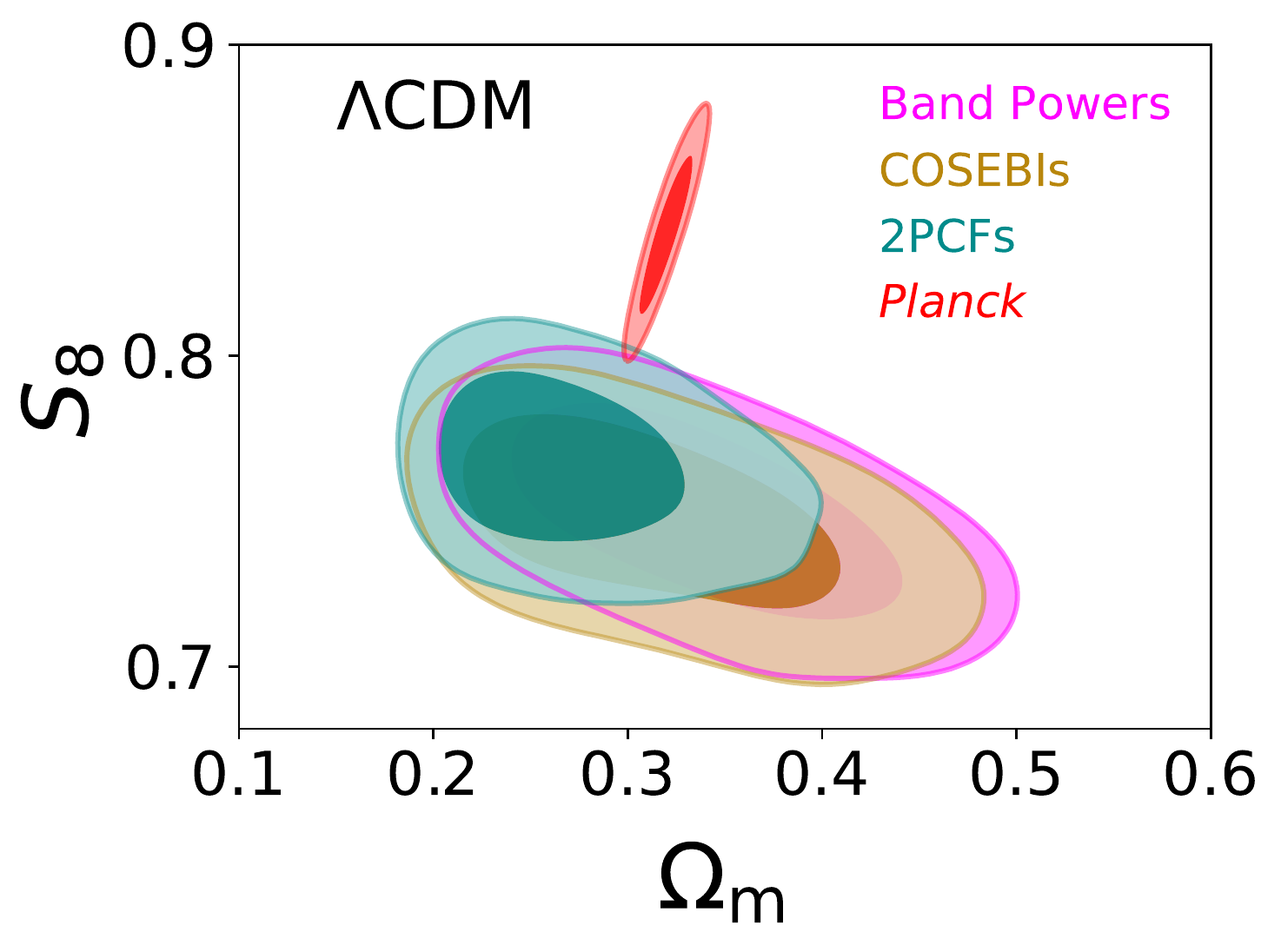}
    \includegraphics[width=0.3\textwidth]{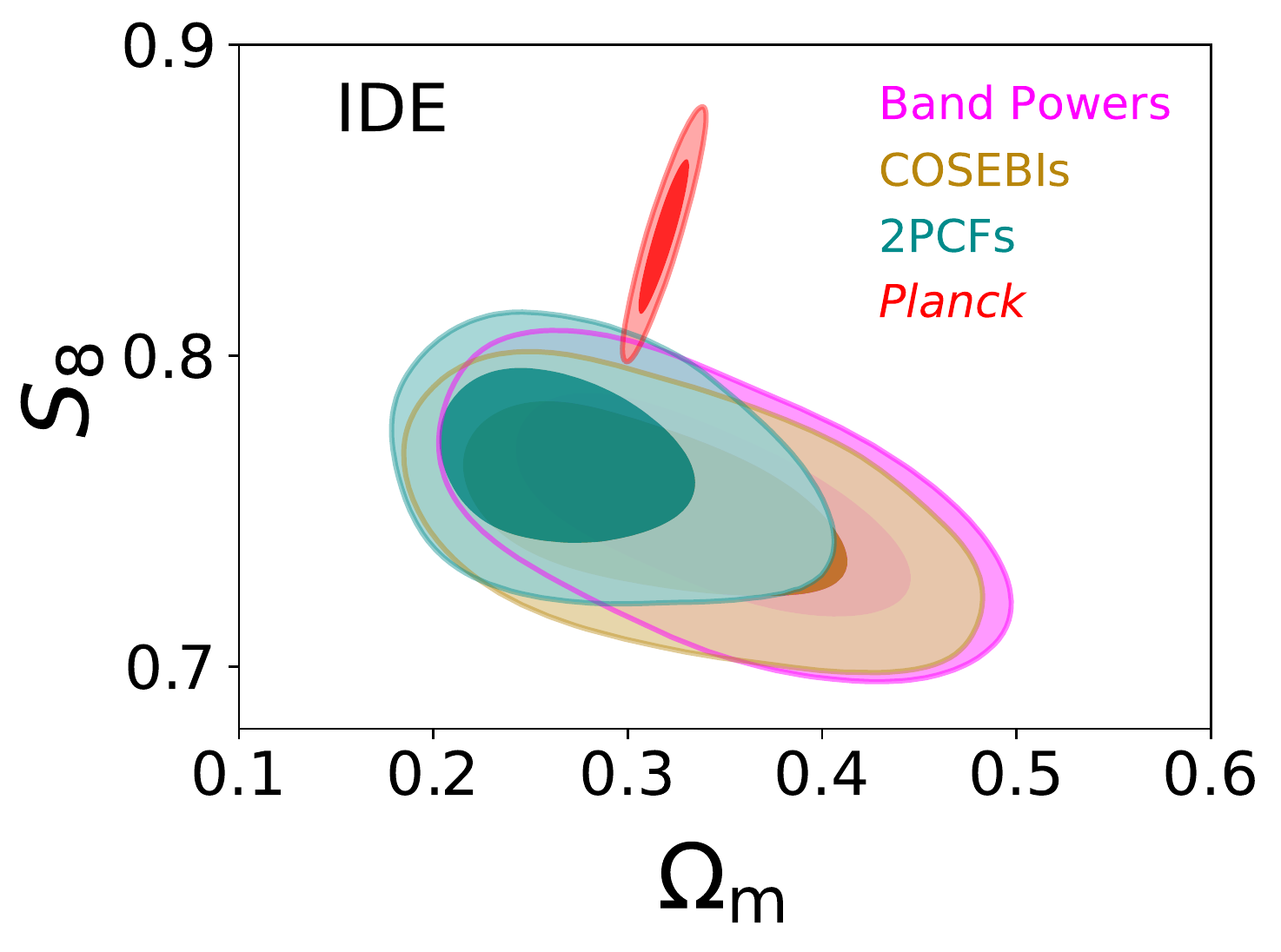}
    \includegraphics[width=0.3\textwidth]{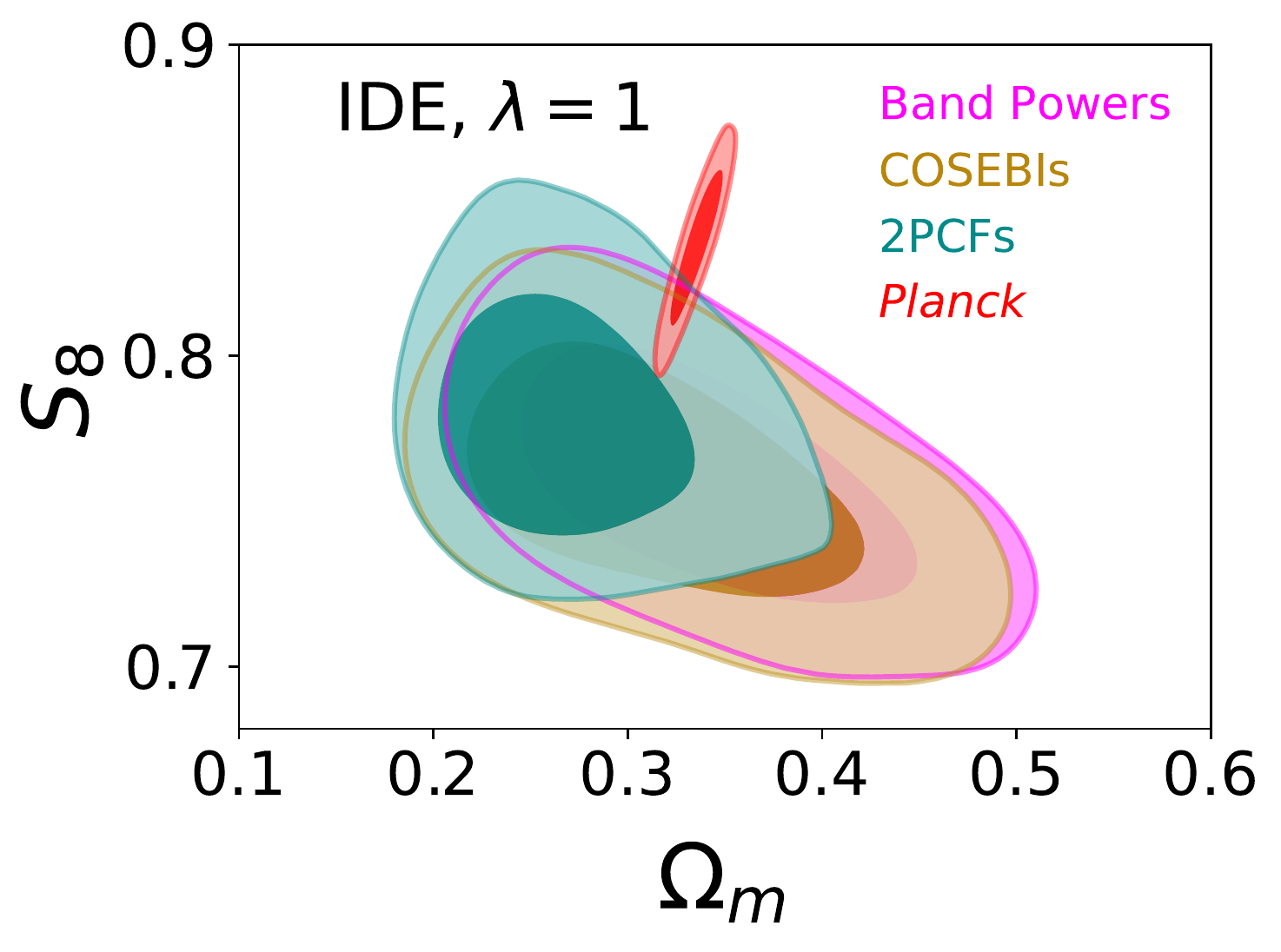}
    \caption{68 and 95 per cent marginalised contours in the $\Omega_{\mathrm{m}}-S_8$ plane. The colour code is the same as in Fig. \ref{fig:constraints_beta}.}
    \label{fig:om_S8}
\end{figure*}

\autoref{tab:posterior} shows the numerical values of the mean and 68 per cent credibility intervals for $\Omega_{\mathrm{m}}, \sigma_8$ and $S_8$, along with $\chi^2$ and log-evidence values, for all cosmic shear summary statistics as well as for \textit{Planck}. \autoref{fig:om_S8} shows contours on the $\Omega_{\mathrm{m}}$-$S_8$ plane for the $\Lambda$CDM and IDE scenarios. The latter is analysed varying both $\beta$ and $\lambda$, as well as setting $\lambda=1$. With this last choice we find an attenuation of the tension up to $\sim$1$\sigma$. In \autoref{tab:posterior} the $\chi^2$ and log-evidence values for $\Lambda$CDM and IDE scenarios (both varying and fixing $\lambda$) are similar across all three summary statistics, hence neither of the two cosmological models is clearly favoured over the other, although the \textit{Planck} data seem to mildly prefer the IDE model over $\Lambda$CDM. Future analyses from Stage IV surveys will have the constraining power to provide stronger model comparison statements. It will be interesting to explore larger prior ranges for $\beta$, as well as different coupling functions, which may lead to stronger alleviation of the $S_8$ tension. For the KiDS-1000 data used in this paper we verified that larger, negative values of $\beta$ do not help alleviate the $S_8$ tension.


\subsection{Forecasts for a \textit{Euclid}-like survey}\label{sec:results_euclid}
\begin{figure*}
    \centering
    \includegraphics[scale=0.14]{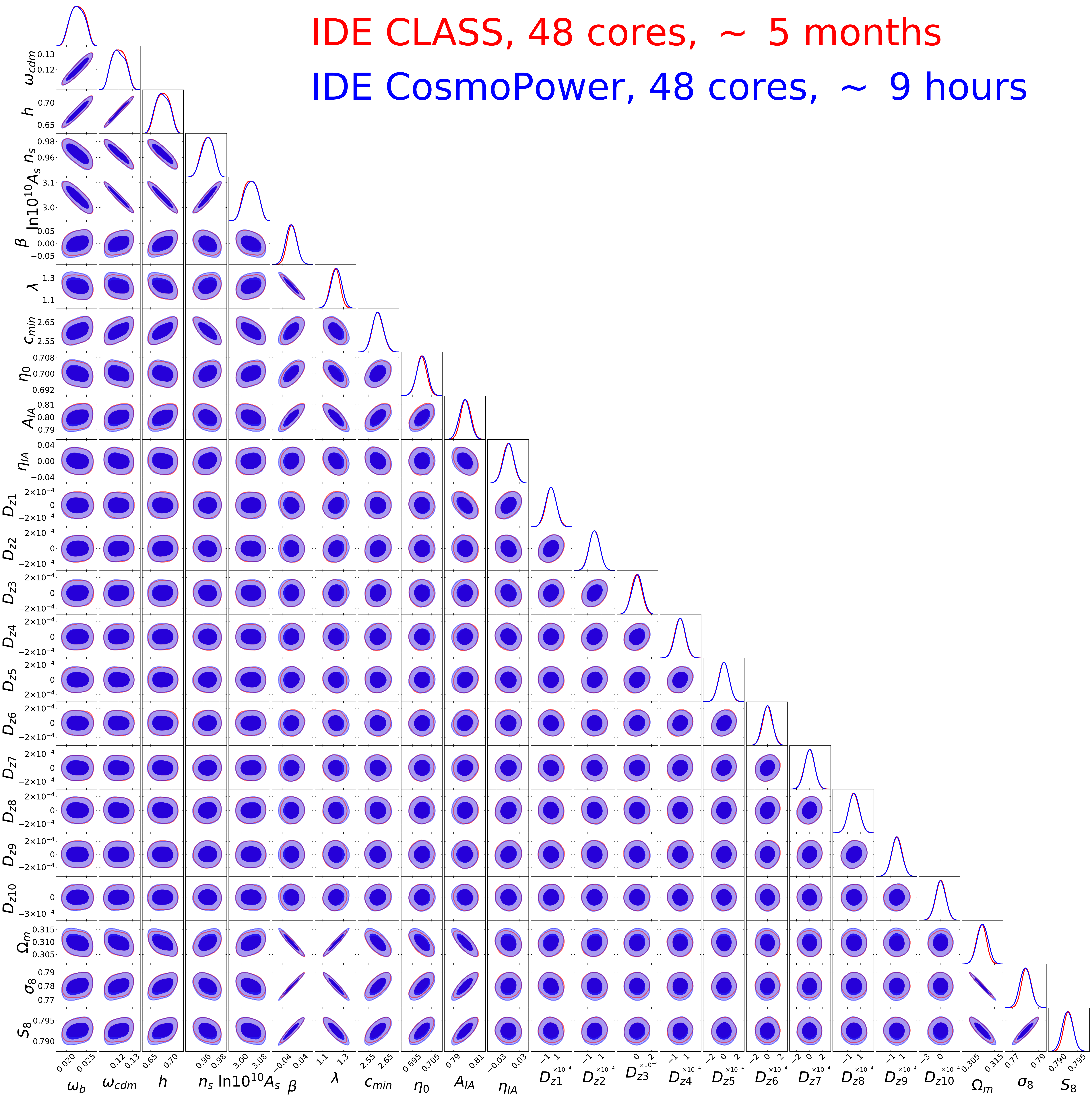}
    \caption{Forecasts for a \textit{Euclid}-like survey. The meaning of each parameter is explained in \citet{SpurioMancini:2021ppk}, whose analysis setup is identical to that considered here, with the sole addition of the interacting dark energy parameters $\beta$ and $\lambda$, introduced in \autoref{sec:IDEmodel}.}
    \label{fig:forecasts_big}
\end{figure*}

In \autoref{fig:forecasts_big} we present forecast contours for a \textit{Euclid}-like Stage IV survey. The simulated configuration is the same presented in \citet{SpurioMancini:2019rxy}, including the prior distributions on cosmological and astrophysical nuisance parameters. For the IDE parameters $\beta$ and $\lambda$, we use prior distributions $\beta \sim \mathcal{U}[-0.5, 0.5]$ and $\lambda \sim \mathcal{U}[0., 2.1]$. We note that the prior on $\lambda$ differs from the one used for the KiDS-1000 data; for future analyses of real data from e.g.~\textit{Euclid} it will be important to consider a uniform prior on log$\, \lambda$ to account for the fact that $\lambda$ is not a dimensionless quantity \citep{Mackay2003}. Here, the goal is to highlight the importance of emulator-based approaches such as the one presented in this paper and based on \textsc{CosmoPower}. With this emulator, we obtained the contours for the \textit{Euclid}-like survey (in \textit{blue} in \autoref{fig:forecasts_big}) in $\sim$ 9 hours running on 48 cores. For comparison, sourcing power spectra from the Boltzmann code \textsc{Class} required a runtime of $\sim$ 5 months on the same hardware configuration (\textit{red} contours in \autoref{fig:forecasts_big}).

We note that this Stage IV survey configuration leads to much stronger constraints on IDE parameters $\beta$ and $\lambda$, namely $\beta = -0.001_{-0.024}^{+0.023}$ and $\lambda = 1.231_{-0.051}^{+0.054}$ (68 per cent contours). We also see that these IDE parameters are degenerate with nuisance parameters $A_{\mathrm{IA}}$ and $\eta_{\mathrm{IA}}$, modelling amplitude and redshift-dependence of the intrinsic alignment signal, respectively, as well as the \textsc{HMcode} parameters $c_{\mathrm{min}}$ and $\eta_0$, describing minimum halo concentration and halo bloating, respectively. These degeneracies highlight the importance of developing accurate prescriptions for nonlinearities and systematics that can guarantee unbiased constraints on dark energy.

\section{Conclusions}
\label{sec:conclusions}

We presented constraints on an interacting dark energy (IDE) model from $\sim 1000$ deg$^2$ cosmic shear measurements from the Kilo-Degree Survey (KiDS-1000). A comparison with \textit{Planck} measurements of the Cosmic Microwave Background (CMB) shows an alleviation up to $\sim$1$\sigma$ of the tension in the parameter $S_8 = \sigma_8 \sqrt{\Omega_{\mathrm{m}}/0.3}$, with respect to the $\sim 3 \sigma$ tension of the $\Lambda$CDM analysis of \citet{Asgari21}. Constraints on the IDE model were obtained taking into account, for the first time, baryonic feedback effects. Given the absence of bespoke nonlinear prescriptions for IDE models, we adopted the $\Lambda$CDM-based nonlinear prescription implemented in the software \textsc{HMcode}. For applications to future surveys, proper nonlinear prescriptions for IDE models will need to be developed. We plan to consider the Elastic Scattering model and the \emph{halo model reaction} framework \citep{Cataneo:2018cic, Bose:2020wch,Troster:2020kai} for this purpose.  
    
In deriving constraints, we used the neural network - based emulator of cosmological power spectra \textsc{CosmoPower} to accelerate the inference pipeline. We highlight the importance of such emulator-based approaches, in particular for applications to Stage IV surveys analyses. To demonstrate this point, we performed a forecast for a Stage IV \textit{Euclid}-like survey for the same IDE model constrained with the KiDS-1000 data. Sourcing power spectra from \textsc{CosmoPower} allowed us to obtain contours in a few hours, while the same contours obtained using a Boltzmann code required a few months of run time. 
    
The emulators trained for this analysis will remain available. For example, following \citet{SpurioMancini:2021ppk}, we emulated the linear matter power spectrum and a nonlinear boost. As new, bespoke nonlinear corrections for IDE models become available, the \textsc{CosmoPower} emulator for the nonlinear boost can be trained on them, while for the linear power spectrum we can reuse the emulator trained for this analysis.

\section*{Acknowledgements}
We thank B. Joachimi and A. Mead for comments on the manuscript, and S. Brieden and T. Tram for useful discussions. We thank the anonymous referee for their valuable review of the paper. ASM is supported by the MSSL STFC Consolidated Grant. AP is a UK Research and Innovation Future Leaders Fellow [grant MR/S016066/1]. This work used facilities provided by the UCL Cosmoparticle Initiative. We acknowledge the use of \textsc{GetDist} \citep{Lewis:1999bs} to obtain corner plots. 
Based on observations made with ESO Telescopes at the La Silla Paranal Observatory under programme IDs 177.A-3016, 177.A-3017, 177.A-3018 and 179.A-2004, and on data products produced by the KiDS consortium. The KiDS production team acknowledges support from: Deutsche Forschungsgemeinschaft, ERC, NOVA and NWO-M grants; Target; the University of Padova, and the University Federico II (Naples).
We used the gold sample of weak lensing and photometric redshift measurements from the fourth data release of the Kilo-Degree Survey \citep{2019A&A...625A...2K, 2020A&A...637A.100W, Hildebrandt21, Giblin21},
referred to as KiDS-1000. Cosmological parameter constraints from KiDS-1000 have been presented in \citet{Asgari21} (cosmic shear), \citet{Heymans_2021} ($3\times2$pt) and \citet{Troster:2020kai} (beyond $\Lambda$CDM), with the methodology presented in \citet{Joachimi21}.
Based on observations obtained with Planck (\url{http://www.esa.int/Planck}), an ESA science mission with instruments and contributions directly funded by ESA Member States, NASA, and Canada.

\section*{Data Availability}

KiDS-1000 data are available at \url{http://kids.strw.leidenuniv.nl/DR4/lensing.php}. The likelihood codes and emulators used in this analysis are shared on the \textsc{CosmoPower} GitHub repository \href{https://github.com/alessiospuriomancini/cosmopower}{\faicon{github}}.



\bibliographystyle{mnras}
\bibliography{references} 



\appendix

\bsp	
\label{lastpage}
\end{document}